\documentstyle[12pt]{article}
\begin{document}
\baselineskip 22pt

\begin{center}
 {\Large UNCONDITIONALLY SECURE QUANTUM BIT COMMITMENT IS POSSIBLE} \\
\vspace*{.4in}

{\Large Horace P. Yuen} \\  {\large Department of Electrical and
Computer Engineering \\ Department of Physics and Astronomy\\
Northwestern University \\ Evanston IL  60208-3118 \\ email:
yuen@ece.northwestern.edu}
\end{center}
\vspace*{.4in}

\begin{abstract}
Bit commitment involves the submission of evidence from one party to
another so that the evidence can be used to confirm a later revealed
bit value by the first party, while the second party cannot determine
the bit value from the evidence alone. It is widely believed that
unconditionally secure quantum bit commitment is impossible due to
quantum entanglement cheating, which is codified in a general
impossibility theorem. In this paper, the scope of this general impossibility proof
is extended and analyzed, and gaps are found.  Three specific
protocols are described for which the entanglement cheating as given
in the impossibility proof fails to work.  One of these protocols,
QBC2, is proved to be unconditionally secure.

\vspace*{.2in}

\noindent PACS \#: 03.67Dd, 03.65Bz
\end{abstract}

\vspace*{.5in}

NOTE

(1) In this v7 of the paper, which is really version 3, the two
    previous versions are subsumed and complete proofs are given for
    all the claims.  The ``history'' of some of the protocols
    discussed can be traced from the previous v1-v6 of this paper.

(2) Many of the points made in this version were mentioned in my
    Capri talk in July 2000.  However, the paper prepared for that
    Proceedings volume, which is available at quant-ph/0009113,
    concentrates on anonymous-key cryptography with only passing
    remarks on bit commitment.

(3) The reader interested only in an unconditionally secure quantum
bit commitment protocol can go directly form section II to section VI.

\newpage

\renewcommand{\thesection}{\Roman{section}}

\tableofcontents

\newpage

\section{\hspace{.2in}Introduction}

Quantum cryptography \cite{bennett}, the study of information security
systems involving quantum effects, has recently been associated almost
exclusively with the cryptographic objective of key distribution.
This is due primarily to the nearly universal acceptance of the general
impossibility of secure quantum bit commitment (QBC), taken to be a
consequence of the Einstein-Podolsky-Rosen (EPR) type entanglement
cheating which rules out QBC and other quantum protocols that have
been proposed for various other cryptographic objectives
\cite{brassard}.  In a bit commitment scheme, one party, Adam,
provides another party, Babe, with a piece of evidence that he has
chosen a bit b (0 or 1) which is committed to her.  Later, Adam would
``open'' the commitment: revealing the bit b to Babe and convincing
her that it is indeed the committed bit with the evidence in her
possession.  The usual concrete example is for Adam to write down the
bit on a piece of paper which is then locked in a safe to be given to
Babe, while keeping for himself the safe key that can be presented
later to open the commitment.  The evidence should be {\em binding},
i.e., Adam should not be able to change it, and hence the bit, after
it is given to Babe.  It should also be {\em concealing}, i.e., Babe
should not be able to tell from it what the bit b is.  Otherwise,
either Adam or Babe would be able to cheat successfully.

In standard cryptography, secure bit commitment is to be achieved
either through a trusted third party or by invoking an unproved
assumption on the complexity of certain computational problem.  By
utilizing quantum effects, various QBC schemes not involving a third
party have been proposed that were supposed to be unconditionally
secure, in the sense that neither Adam nor Babe can cheat with any
significant probability of success as a matter of physical laws.  In
1995-1996, a general proof on the impossibility of unconditionally
secure QBC and the insecurity of previously proposed protocols were
described \cite{mayers1}-\cite{lo}.  Henceforth, it has been accepted
that secure QBC and related objectives are impossible as a matter of
principle \cite{lo2}-\cite{brassard2}.

There is basically just one impossibility proof, which gives the EPR
attacks for the cases of equal and unequal density operators that Babe
has for the two different bit values. The proof shows that if Babe's
successful cheating probability $P^B_c$ is close to the value 1/2,
which is obtainable from pure
guessing of the bit value, then Adam's successful cheating probability $P^A_c$ is
close to the perfect value 1.  This result is stronger than the mere impossibility of unconditional security, namely that it is
impossible to have both $P^B_c \sim 1/2$ and $P^A_c \sim 0$. Since
there is no known characterization of all possible QBC protocols,
logically there can really be no general impossibility proof, strong
or not, even if
it were indeed impossible to have an unconditionally secure QBC
protocol. This problem of scope of the impossibility proof can be seen
from the following simple example.

Suppose Adam commits a state $| \phi \rangle$ of a single qubit
(two-dimensional quantum state space) for the bit value 0 and $| \phi'
\rangle$ for 1. Adam opens by declaring the bit value, and Babe verifies by measuring the corresponding
projection, $| \phi \rangle \langle \phi |$ or $| \phi' \rangle
\langle \phi' |$. It is intuitively clear, but will be
formalized as local state invariance in this paper, that Adam can
launch no effective EPR cheat.  Of course, it is true in this case
that if $P^B_c \sim 1/2$ then $| \langle \phi | \phi' \rangle |^2 \sim
1$, so $P^A_c \sim 1$ simply by declaring the bit value 1 even when $| \phi
\rangle$ is committed.  However, it is a priori possible for a
protocol to have the property that $P^B_c \sim 1/2$ while Adam cannot
form any effective cheating entanglement as in this example but with
$P^A_c \sim 0$.  To have a general impossibility proof,
one has to show that this property cannot be obtained in any QBC
protocol or that any unconditionally secure QBC protocol would contradict some known
principle.  The mere absence of counterexamples does not constitute a
proof.

The general questions of scope of the impossibility proof will be
addressed specifically in Section IV.  Three QBC schemes not covered
by the impossibility proof will be described in
Sections V-VII, although only one of them, QBC2 in Section VI, is
proved to be unconditionally secure in this papper. The results are
developed within nonrelativistic quantum mechanics, unrelated to
relativistic protocols \cite{kent} or cheat-sensitive protocols
\cite{hardy}. The essential point is that the flow of classical
information between Adam and Babe in the protocol is crucial to the
possible operations they can carry out, hence fundamentally affecting
the security level of the scheme.  In the impossibility proof, it is
basically assumed that both Adam and Babe possess full information at
each stage of the protocol, an unwarranted assumption.

In Section II the impossibility proof will be reviewed.  Since the
issues involved in quantum cryptography, or classical cryptography for
that matter, are often subtle, it is the policy of this paper to give
complete proofs for its claims.  Thus, the gap between the
quantitative impossibility claim and the result available in the
literature will be filled. An in-principle insecure protocol QBC0 is
also described that underlies QBC1 and QBC3.  In Section III, the
impossibility proof in the original formulation is extended to cover
the situation in which Babe applies a superoperator transformation to
Adam's committed state before perfect verification. Another insecure
protocol QBC01, related to QBC0, is described as an illustration. The reader who just
wants to see an unconditionally secure QBC protocol can go directly
from Section II to Section VI.  Note that the results in this paper
are valid in infinite-dimensional spaces.  Also, the same index
symbols $i,j$, etc., may denote different quantities in different sections.

\newpage

\section{\hspace{.2in}The Impossibility Proof}

In this Section we review the standard formulation of the
impossibility proof and then recast it in a form more suitable
for quantitative analysis and extension, and describe a protocol QBC0.  The
development of this section will be used in the rest of the paper.

According to the impossibility proof, Adam would generate $|\Phi _0
\rangle$ or $|\Phi _1 \rangle$ depending on b = 0 or 1,
\begin{equation}
|\Phi _0 \rangle = \sum_i \sqrt{p_i} | e_i \rangle | \phi _i \rangle,
\label{ent1}
\end{equation}
\begin{equation}
 |\Phi _1 \rangle = \sum_i \sqrt{p'_i} | e'_i \rangle | \phi ' _i
  \rangle
\label{ent2}
\end{equation}
where the states $\{ | \phi_i \rangle \}$ and $\{ | \phi'_i \rangle
\}$ in ${\mathcal H}^B$ are openly known, $i \in \{1, \ldots, M \}$,
$\{ p_i \}$ and $\{ p'_i \}$ are known probabilities, while $\{ | e_i
\rangle \}$ and $\{ | e'_i \rangle \}$ are two complete orthonormal
sets in ${\mathcal H}^A$. All Dirac kets are normalized in this
paper. Adam sends Babe ${\mathcal H}^B$ while keeping ${\mathcal H}^A$
to himself.  He opens by measuring the basis $\{ | e_i \rangle \}$ or $\{
| e'_i \rangle \}$ in $\mathcal{H}^A$ according to his committed state
$| \Phi_0 \rangle$ or $| \Phi_1 \rangle$, resulting in a specific $|
\phi_i \rangle$ or $| {\phi'}_i \rangle$ on $\mathcal{H}^B$, and
telling Babe which $i$ he has obtained.  Babe verifies by measuring
the corresponding projector and will obtain the value 1 (yes) with
probability 1. Adam can, as was argued, switch between $|\Phi _0
\rangle$ and $|\Phi _1 \rangle$ by operation on ${\mathcal H}^A$
alone, and thus alter the evidence to suit his choice of b before
opening the commitment.  In the case $\rho^B_0 \equiv {\rm tr}_A |\Phi
_0 \rangle \langle \Phi _0 | = \rho ^B _1 \equiv {\rm tr}_A |\Phi _1
\rangle \langle \Phi _1|$, the switching operation is to be obtained
by using the so-called ``Schmidt decomposition
\cite{schmidt_decomp},'' the expansion of $|\Phi _0 \rangle$ and
$|\Phi _1 \rangle$ in terms of the eigenstates $|\hat{\phi}_k \rangle$
of $\rho^B_0 = \rho^B_1$ with eigenvalues $\lambda_k$ and the eigenstates $|\hat{e}_k \rangle$ and
$|\hat{e}'_k \rangle$ of $\rho^A_0$ and $\rho^A_1$,
\begin{equation}
|\Phi _0 \rangle = \sum _k \sqrt{\lambda_k} |\hat{e}_k \rangle
|\hat{\phi}_k \rangle, \hspace*{.2in} |\Phi _1 \rangle = \sum _k
\sqrt{\lambda_k} |\hat{e}'_k \rangle |\hat{\phi}_k \rangle
\end{equation}
By applying a unitary $U^A$ that brings $\{ |\hat{e}_k \rangle \}$ to
$\{ |\hat{e}'_k \rangle \}$, Adam can select between $|\Phi_0 \rangle$
or $|\Phi_1 \rangle$ any time before he opens the commitment but after
he supposedly commits.  When $\rho_0^B$ and $\rho_1^B$ are not equal
but close, it was shown that one may transform $| \Phi _0 \rangle$ by
an $U^A$ to a $| \tilde{\Phi} _0 \rangle$ with $|\langle \Phi _1 |
\tilde{\Phi} _0 \rangle |$ as close to 1 as $\rho ^B_0$ is close to
$\rho ^B_1$ according to the fidelity F chosen, and thus the state $|
\tilde{\Phi} _0 \rangle$ would serve as the effective EPR cheat.

In addition to the above quantitative relations, the gist of the
impossibility proof is supposed to lie in its generality -- that any
QBC protocol could be fitted into its formulation, as a consequence of
various arguments advanced in \cite{mayers1}-\cite{brassard2}. Among
other reasons to be elaborated in Section IV, it appeared to the
present author from his development of a new cryptographic tool,
anonymous quantum key technique \cite{yuen_capri}, that the
impossibility proof is not sufficiently general. First of all, there
is {\em no} need for Adam to entangle anything in an honest protocol.
When Adam picks b=0, he can just send Babe a state $| \phi_i \rangle$
with probability $p_i$.  When he picks b=1, he sends $| \phi'_i
\rangle$ with probability $p'_i$.  If the anonymous key technique is
employed, $| \phi_i \rangle$ and $| \phi'_i \rangle$ are to be
obtained from applying $U_i$ or $U'_i$ from some fixed openly known
set of unitary operators $\{ U_i \}$ and $\{ U'_i \}$ on ${\mathcal
H}^B$ by Adam to the states $| \psi \rangle$ sent to him by Babe and
known only to her. As a consequence, Adam would not be able to
determine the cheating unitary transformation $U^A$ as in protocol
QBC1, to be described in Section V after the impossibility proof is
first analyzed generally.

In a QBC protocol, the $\{ | \phi_i \rangle \}$ and $\{ | \phi'_i \rangle \}$
are chosen so that they are concealing as evidence, i.e. Babe cannot
reliably distinguish them in optimum binary hypothesis testing
\cite{helstrom}. (The role of quantum detection theory in QBC together
with some new results used in this paper are elaborated in Appendix A).
They would also be binding if Adam is honest and sends them as they
are above, which he could not change after Babe receives them.  Babe
can always guess the bit with a probability of success $P^B_c = 1/2$,
while Adam should not be able to change a committed bit at all.
However, it is meaningful and common to grant {\em unconditional
security} when the best $\bar{P}^B_c$ Babe can achieve is arbitrarily
close to 1/2 and Adam's best probability of successfully changing a
committed bit $\bar{P}^A_c$ is arbitrarily close to zero even when both parties have perfect technology and unlimited resources
including unlimited computational power \cite{mayers}.  To facilitate the
quantitative analysis of these performance measures, the impossibility
proof would first be reformulated.

Before proceeding, note the following basic property of entanglement
important in QBC.

\noindent {\em Theorem (Local State Invariance):}  Let $\rho^{AB}$ be
a state on ${\cal H}^A \otimes {\cal H}^B$ with marginal states $\rho
^A \equiv tr _B \rho^{AB} , \rho ^B$.  The individual or combined
effects of any state transformation and quantum measurement (averaged
over the measurement results) on ${\cal H}^A$ alone leaves $\rho ^B$
invariant.

See Appendix B for a proof and a discussion of its
role in the impossibility of superluminal communication via quantum entanglement.

As a consequence of this theorem, Adam cannot cheat by changing the
$\rho^B_0 \neq \rho^B_1$ case to the $\rho^B_0 = \rho^B_1$ case
whatever the $\rho$'s are.  In particular, a single pure state as in
the example of Section I cannot be changed.

The operation of unitary transformation with subsequent measurement of
an orthonormal basis is equivalent to the mere measurement of another
orthonormal basis $\{ |\tilde{e}_i \rangle \}$ on the system.  Thus,
the net cheating operation can be described by writing
\begin{equation}
| \Phi_0 \rangle = \sum_i \sqrt{\tilde{p}_i} | \tilde{e}_i \rangle |
  \tilde{\phi}_i \rangle,
\end{equation}
\begin{equation}
\sqrt{\tilde{p}_i}|\tilde{\phi}_i \rangle \equiv \sum_j \sqrt{p_j}
V_{ji} |\phi_j \rangle
\end{equation}
for a unitary matrix V defined by $| e_i \rangle = \sum_j V_{ij} |
\tilde{e}_j \rangle$, and then measuring $|\tilde{e}_i\rangle$. For
convenience, we may still in the rest of the paper refer to the
cheating operation as a $U^A$ transformation described at the
beginning of this Section.  Local state invariance is a property complementary to the
fact that the $| \tilde{\phi}_i \rangle$ obtainable by operation on
${\mathcal H}^A$ alone are some proper linear combinations of the $|
\phi_i \rangle$ from (5).  The quantitative expression for $P^A_c$ can
now be given.  If Babe verifies the individual $| \phi'_i \rangle$,
the Adam's successful cheating probability is
\begin{equation}
P^A_c = \sum_i \tilde{p}_i | \langle \tilde{\phi}_i | \phi'_i \rangle
|^2.
\end{equation}

In general, the optimal cheating probability $\bar{P}^B_c$ for Babe is
given by the probability of correct decision for optimally discriminating between two density operators $\rho^B
_0$ and $\rho^B _1$ by any quantum measurement.  From (A4) with $p_0 =
1/2$,
\begin{equation}
\bar{P}^B_c = \frac{1}{4} (2 + \| \rho^B _0 - \rho^B _1 \| _1)
\end{equation}
where $\| \cdot \| _1$ is the trace norm, $\| \tau \| _1 \equiv
tr(\tau ^{\dagger} \tau )^{1/2}$, for a trace-class operator $\tau$
\cite{schatten}.  In terms of a security parameter $n$ that can be
made arbitrarily large ,the statement of unconditional security (US) can be
quantitatively expressed as
\begin{equation}
{\rm (US)} \qquad \qquad \lim _n \bar{P} ^B _c = \frac {1} {2} \quad {\rm
and} \quad \lim _n \bar{P} ^A _c = 0.
\end{equation}
Condition (US) is equivalent to the statement that for any $\epsilon > 0$, there exists an
$n_0$ such that for all $n > n_0$, $\bar{P}^B_c - \frac{1}{2} <
\epsilon$ and $\bar{P}^A_c < \epsilon$, i.e. $\bar{P}^B_c -
\frac{1}{2}$ and $\bar{P}^A_c$ can both be made arbitrarily small for
sufficiently large $n$.  The impossibility proof claims a lot more
than the mere impossibility of (US), it asserts \cite{mayers} the
following statement (IP):
\begin{equation}
{\rm (IP)} \qquad \lim _n \bar{P} ^B _c = \frac {1} {2} \quad
\Rightarrow \quad \lim _n \bar{P} ^A _c = 1.
\end{equation}
In the $\rho ^B _0 = \rho ^B _1$ case, the EPR cheat shows that
$\bar{P}^B_c=\frac{1}{2}$ implies $\bar{P}^A_c=1$.  Thus (IP)
generalizes it to the assertion that the function
$\bar{P}^A_c(\bar{P}^B_c)$, obtained by varying $n$, is {\em
continuous} from above at $\bar{P}^B_c = \frac{1}{2}$. Note the
considerable difference between the truth of (IP) and the
much weaker statement that (US) is impossible. In the middle ground
that $\lim_n \bar{P}^B_c = \frac{1}{2}$ implies just $0 < \lim_n
\bar{P}^A_c < 1$, the protocol would be concealing for Babe and
cheat-sensitive for Adam.

The key observation \cite{mayers1}-\cite{mayers} in the proof of (IP)
is the use of Uhlmann's theorem, that there exist purifications $|
\Phi_0 \rangle$ and $| \Phi_1 \rangle$ of any given $\rho_0$ and
$\rho_1$ such that $| \langle \Phi_0 | \Phi_1 \rangle |^2$ attains the
maximum possible value given by $F(\rho_0,\rho_1) \equiv \left| {\rm
tr} \sqrt{\sqrt{\rho_0} \rho_1 \sqrt{\rho_0}} \right|^2$.  The
conclusion is drawn, without supporting details, that if
$F(\rho^B_0,\rho^B_1)$ is close to 1, then so is $\bar{P}^A_c$.  This
conclusion can be related to (IP) via the bound \cite{fuchs}
\begin{equation}
2 [1-\sqrt{F(\rho_0,\rho_1)}] \le \| \rho_0 - \rho_1 \| _1.
\label{trnorm_bound}
\end{equation}
as follows. Let $\| \rho^B_0 - \rho^B_1 \| _1 \le \epsilon$, so that
$F(\rho^B_0,\rho^B_1) \ge (1-\frac{\epsilon}{2})^2$ from
(\ref{trnorm_bound}).  From Uhlmann's theorem, choose $| \Phi_0
\rangle$ and $| \Phi_1 \rangle$ of (1)-(2) to be the purifications
that achieve the maximum $F(\rho^B_0,\rho^B_1)$ so that $| \langle
\Phi_0 | \Phi_1 \rangle | \ge 1 - \frac{\epsilon}{2}$.  The cheating
operation on $|\Phi_0\rangle$ is given by (5), and Adam's successful
cheating probability is given by the following \\
\noindent{\em{Lemma 1}}: For probabilities $\alpha_i$ and complex
numbers $\lambda_i$,
\begin{equation}
\sum_i \alpha_i | \lambda_i |^2 \ge \left| \sum_i \alpha_i \lambda_i
\right|^2
\label{lemma1}
\end{equation}
(the sums can be over infinite sets).  \\
\noindent{\em{Proof}}: When $\lambda_i$ are real, (\ref{lemma1}) follows from
Jensen's inequality \cite{polya} and the concavity of the function $x
\mapsto x^2$.  The complex case follows by expanding each $\lambda_i$
into real and imaginary parts. $\Box$

\noindent{Since $\langle \tilde{\Phi}_0 | \Phi_1 \rangle = \sum_i \sqrt{p'_i
\tilde{p}_i} \langle \tilde{\phi}_i | \phi'_i \rangle$, it follows
from (\ref{lemma1}) with $\lambda_i =\sqrt{\tilde{p}_i/p'_i} \langle
\tilde{\phi}_i | \phi'_i \rangle$ (no need to include the $p'_i=0$
terms) and $\alpha_i = p'_i$ that $\bar{P}^A_c \ge 1-\epsilon$
whenever $\bar{P}^B_c \le \frac{1}{2} + \frac{\epsilon}{4}$.  Thus, the statement (IP) is
proved. In particular, one has the convergence rate
\begin{equation}
\bar{P}^B_c=\frac{1}{2}+O\left(\frac{1}{n}\right) \Rightarrow
\bar{P}^A_c=1-O\left(\frac{1}{n}\right).
\end{equation}

As an illustration, consider the following protocol, in which
hashing via the parity function is used to make $\rho^B_0$ close to
$\rho^B_1$ in a sequence generalization of the example in Section~I.

{\em PROTOCOL} QBC0:

(i) Adam sends Babe a sequence of $n$ qubits, each is either one of
$\{ | \phi \rangle, | \phi' \rangle \}$, such that an even number of
$| \phi' \rangle$ corresponds to b = 0 and an odd number to b = 1,
with probability $1/2^{n-1}$ for each sequence of either parity.

(ii) Adam opens the commitment by revealing the sequence of $n$
states.  Babe verifies by measuring the corresponding projection on
each qubit to see that the entire sequence is correct.

To show that this scheme can be made concealing, note that $\rho^B_0 -
\rho^B_1$ factorizes into products of individual qubit parts as
follows.  Let ${\bf j}=(j_1,\ldots,j_n) \in \{0,1\}^n$,
$P_{l0}=|\phi\rangle\langle\phi|$,
$P_{l1}=|\phi'\rangle\langle\phi'|$, $l \in \{1,\ldots,n\}$. Let $\Lambda_0 = \{ {\bf j} |
\bigoplus^n_{l=1} j_l = 0 \}$, $\Lambda_1 = \{ {\bf j} |
\bigoplus^n_{l=1} j_l = 1 \}$ be the even and odd parity $n$-bit sets.
Then
\begin{equation}
\rho^B_i = \frac{1}{2^{n-1}} \sum_{{\bf j} \in \Lambda_i}
\, \bigotimes^n_{l=1} P_{lj_l},\qquad i \in \{0,1 \}
\end{equation}
and so
\begin{equation}
\rho^B_0 - \rho^B_1 = \frac{1}{2^{n-1}} \bigotimes^n_{l=1}
(P_{l0}-P_{l1}).
\end{equation}
Thus, Babe's optimum quantum decision reduces to optimally deciding
between $| \phi \rangle$ and $| \phi' \rangle$ for each qubit
individually and then seeing whether there is an even or odd number of
$|\phi'\rangle$'s.  The optimum error probability $p_e$ for each qubit
is given in (A5), and the optimum error probability $\bar{P}^B_c$ of
correct bit decision on the sequence is, from the even and odd binomial
sums (cf. Appendix C),
\begin{equation}
\bar{P}^B_c=\frac{1}{2}+\frac{1}{2}(1-2p_e)^n.
\end{equation}
Thus, $\bar{P}^B_c$ is close to $\frac{1}{2}$ exponentially in $n$
independently of $\frac{1}{2} \ge p_e > 0$.  However, Adam can now cheat by forming
entanglement as in (1)-(2), with $\bar{P}^A_c$ exponentially close to
1 in accordance with (IP).

\newpage

\section{\hspace{.2in}An Extension of the Impossibility Proof}

In this Section, a protocol QBC01 will be described in which Babe
introduces a lossy transformation on Adam's committed state while
still being able to verify perfectly.  While it may be argued that
such transformation cannot succeed in obtaining a secure protocol on
qualitative grounds, it may also be argued otherwise.  Specifically,
the coherence of the states (1)-(2) can be deliberately destroyed by
Babe with such a CP map, reducing the entangled states to incoherent
superpositions in her observation space.  It turns out that if she does that, which she can
emphatically do, the resulting condition on the number $n$ of modes would not fit with
the other requirements of the protocol.  Indeed, the impossibility
proof will be extended to cover all such possibilities of Babe
introducing a CP-map transformation.

The following protocol is closely related to QBC0.

{\em PROTOCOL} QBC01.

(i) Adam sends Babe a sequence of $n$ states $|\alpha_l\rangle \in
{\mathcal H}^B_l$, ${\mathcal H}^B = \bigotimes_l {\mathcal H}^B_l$,
each $|\alpha_l\rangle$ being either one of two coherent states $\{ |
\alpha \rangle, |\alpha' \rangle\}$, such that an even number of
$|\alpha'\rangle$ corresponds to b = 0 and an odd number to b = 1,
with probability $1/2^{n-1}$ for each sequence of either parity.

(ii) Babe splits each state $|\alpha_l \rangle$ to
$|\sqrt{\eta}\alpha_l\rangle$ on ${\mathcal H}^B_l$, $\eta < 1$.

(iii) Adam opens the commitment by revealing the sequence of $n$
states.  Babe verifies by measuring the corresponding projection on
each $| \sqrt{\eta} \alpha_l \rangle$ to see that the entire sequence
is correct.

The cheating transformation on this protocol would produce from (5) a
superposition of coherent states with large energy difference when
$|\alpha - \alpha'| >> 1$.  As explained in version 2 of this paper
(v4-v6), such superpositions are supersensitive to loss
\cite{caldeira}-\cite{yuen_highrate}, thus offering the possibility that a
lossy transformation, which would not afect perfect verification on
coherent states, would destroy the necessary entanglement for Adam to
cheat successfully.  However, in this multimode situation, in order to
destroy the coherence one needs to have a loss of one photon {\em per
mode}, not just one photon, and the protocol cannot be made secure.
Indeed, this condition on the destruction of coherence is what makes
fault-tolerant quantum computing in the presence of loss possible.

We now give the impossibility proof that defeats such a maneuver by
Babe. Let ${\mathcal J}_B$ be any completely positve (CP) map
(superoperator) on density operators introduced by Babe.  Let
$X^B_{i1}$ be the measurement operator that perfectly verifies the b =
1 case given $i$, i.e. $X^B_{i1}$ is the $\Pi_1$ operator part of a
POM for the ``1'' or ``not 1'' decision in quantum hypothesis
testing as described in Appendix A, with perfect verification
corresponding to the condition
\begin{equation}
{\rm tr}X^B_{i1}{\mathcal J}_B|\phi'_i\rangle\langle\phi'_i|=1.
\end{equation}
The $P^A_c$ then becomes
\begin{equation}
P^A_c = \sum_i \tilde{p}_i {\rm tr} X^B_{i1} {\mathcal
J}_B|\tilde{\phi}_i\rangle\langle\tilde{\phi}_i|.
\end{equation}
The following lemma and all other results in this paper are valid in
infinite-dimensional spaces.  \\
\noindent{\em{Lemma 2}} \cite{schatten}:  For any bounded operator $X$
and any trace-class operator $\tau$,
\begin{equation}
|{\rm tr} X\tau| \le \| X \| \|\tau\|_1,
\label{lemma2}
\end{equation}
where $\| \cdot \|$ is the ordinary operator norm.

Since $\| X^B_{i1} \| \le 1$, from (\ref{lemma2}) we get
\begin{equation}
P^A_c \ge 1 - \sum_i \tilde{p}_i \left\| {\mathcal{J}}_B (|
\tilde{\phi}_i \rangle \langle \tilde{\phi}_i | - | \phi^{\prime}_i
\rangle \langle \phi^{\prime}_i | ) \right\| _1.
\end{equation}
From the original $P^A_c = \sum_i \tilde{p}_i |\langle \tilde{\phi}_i
| \phi^{\prime}_i \rangle |^2 \ge 1-\epsilon$ for $\| \rho^B_0 -
\rho^B_1 \| \le \epsilon$ proved in Section II, one obtains, by relating inner
product and trace norm for pure states as in (A4)-(A5),
\begin{equation}
\sum_i \tilde{p}_i \left \| | \tilde{\phi}_i \rangle \langle
\tilde{\phi}_i | - | \phi^{\prime}_i \rangle \langle \phi^{\prime}_i |
\right \| ^2 _1 \le 4 \epsilon.
\end{equation}
The following theorem is actually valid for any positive
trace-preserving map $\mathcal{J}$.  \\
\noindent {\em Theorem} \cite{ruskai}:
\begin{equation}
\| {\mathcal{J}} (\rho_0 - \rho_1) \| _1 \le \| \rho_0 - \rho_1 \| _1.
\end{equation}
From (20) and (21), $\sum_i \tilde{p}_i \left \| {\mathcal{J}}_B
\left( | \tilde{\phi}_i \rangle \langle \tilde{\phi}_i | - |
\phi^{\prime}_i \rangle \langle \phi^{\prime}_i | \right ) \right \|
^2 _1 \le 4 \epsilon$ and, using (\ref{lemma1}),
\begin{equation}
\sum_i \tilde{p}_i \left \|  {\mathcal J}_B \left( \tilde{\phi}_i
\rangle \langle \tilde{\phi}_i | - | \phi^{\prime}_i \rangle \langle
\phi^{\prime}_i | \right) \right \| _1 \le 2 \sqrt{\epsilon}.
\end{equation}
Putting (22) into (19) yields $P^A_c \ge 1-2\sqrt{\epsilon}$,
completing the proof of (IP). It appears that the use of the trace norm cannot be
avoided here, in contrast to the ${\mathcal{J}}_B = I^B$ case, which is
responsible for the weakening of the $\bar{P}^A_c$ convergence rate
from $1-O\left(\frac{1}{n}\right)$ to
$1-O\left(\frac{1}{\sqrt{n}}\right)$.

The perfect verification condition (16), preserved in protocol QBC01, is
not necessary for a secure QBC protocol.  This point and the
entanglement destruction strategy of protocol QBC01 will be exploited
in protocol QBC3 of Section VII.  These possibilities also suggest
that it is now appropriate to examine the assumptions underlying the
impossibility proof.

\newpage

\section{\hspace{.2in}The Limited Scope of the Impossibility Proof}

The generality of the scope of the impossibility proof is analyzed in
this section on general grounds.  This is an important issue because
unconditionally secure bit commitment is too useful to give up easily,
and the available impossibility proof has many weaknesses that can be
exploited for secure QBC protocols.  At the very least, one may hope
that hidden assumptions, perhaps practically valid, may be revealed.
Indeed one such assumption is that the quantum and classical
communications involved are over a perfect channel, which should be
considered different from the assumption that the parties have perfect technology.  This
is a good assumption for some situations, but not others such as
long-distance fiber-optic communications.  Another example in which
this assumption is not valid involves satellite-to-satellite optical
communications where the receivers' fields of view have to be opened up,
perhaps because the signals are deliberately spread, so that the sun's
background contributes a significant amount of noise.  In both of
these cases, one can stretch the meaning of ``perfect technology'' to
say that no unavoidable classical disturbance needs to be present --
say, by throwing the sun to another galaxy. (And what about the cosmic
background radiation?) But then the relevance of
such results to reality is quite questionable. In this paper, a
perfect channel is granted. Since it is widely
believed that there is a complete impossibility proof in such a case, I would try to show otherwise independently of the
protocols of the paper.

The major problem is, of course, to decide whether the
formulation given in \cite{mayers1}-\cite{brassard2} is sufficiently
broad to include {\em all} possible QBC protocols.  Typically, one
proves general impossibility by showing that any concretely suggested
possibility would lead to a contradiction. The simplest example
is that the possibility of superluminal communication via quantum
entanglement would contradict local state invariance
(cf. Appendix B).  Another example
would be the quantum no-clone theorem, where cloning contradicts
unitarity on a sufficiently large Hilbert space \cite{yuen1} as well
as quantum detection theory (cf. Appendix A).  In von Neumann's famous no-hidden-variable theorem
\cite{neumann}, a contradiction is derived from what he considered to be
the requirements for a hidden-variable theory.  Perhaps more
significant and illuminating is the impossibility proof of certain
geometric constructions by straightedge and compass developed in the
first half of the nineteenth century, in which any such construction is characterized by
the membership of a certain number lying in a tower of quadratic
extension fields \cite{jacobson}.  This example is significant because
it is nontrivial to capture enough of the essence of {\em any}
straightedge-and-compass construction to be able to produce a
mathematical contradiction when the construction is impossible.  Thus,
for a general impossibility proof of unconditionally secure QBC, one
would expect that the general essence of any such protocol would be
extracted to yield a contradiction.  Clearly the impossibility proof
does not do that, but rather relies on the claim that any possible QBC
protocol can be reduced to its formulation. It is not a priori
impossible to exhaustively describe and classify all operations of a
certain kind, say, in quantum key distribution one typically
characterizes all possible attacks Eve can launch.  However, it is
much more difficult to characterize all possible protocols than all
possible attacks for any cryptographic objective because an arbitrary
interactive flow of information between users is possible in a
protocol.  Indeed, no characterization of all protocols for a specific
objective is known in standard (classical) cryptography. The scope problems of the impossibility proof are numbered
as follows.

(1) One justification for the all-encompassing nature of the
    formulation is that Adam is proceeding exactly as if he were
    honest, except right before opening, in carrying out his EPR cheat.
    This is not true because there is no need for him to entangle
    anything in an honest protocol.  He can just pick a
    $|\phi_i\rangle$ or $|\phi^{\prime}_i\rangle$ and send it.

(2) Because of this, it is not clear why Adam must be able to form the
    entanglement he needs for any possible protocol.

(3) Furthermore, it is not clear why Adam must be able to determine
    the cheating transformation, even apart from complexity questions,
    for any possible protocol. Protocol QBC1 of
    section V provides a direct challenge in this situation, while
    protocol QBC2 of Section VII can also be considered to pose this problem.

(4) The formulation postpones any measurement to the end of the
    commitment phase and claims that it entails no loss of
    generality.  But why
    wouldn't it affect the quantitative cheating probabilities?
    Protocol QBC2 provides an example in which the timing of the
    measurement has substantial consequence.

(5) The density operators $\rho^B_0$ and $\rho^B_1$ for Babe are not
    necessarily the marginal states obtained from the states generated
    by Adam because of Babe's possible lack of information, a situation that is
    built into the protocol.  Thus, Adam's EPR cheat may not
    correspond to the $\rho^B_0 \sim \rho^B_1$ case.  An example is
    provided by QBC2.

(6) It is clearly possible to avoid EPR cheats, as in the
    example described in Section I. While (IP) holds in this case, it holds {\rm not} because of EPR cheats.  The question is:
    why is it that an EPR-cheat-free protocol necessarily cannot
    satisfy (US)?  Protocol QBC2 is an explicit example that (US) is
    possible in such a protocol.

(7) It is not clear why perfect verification is necessary, the only
    performance measures here being the cheating probabilities.  This
    freedom in a QBC protocol is exploited in QBC3.

(8) It is not clear why Babe is necessarily unable to destroy Adam's
    entanglement by her action alone.  Despite the failure of QBC01 of
    Section III, this possibility is manifested in protocol QBC3.

The list could be continued. Note that the {\em burden} is on the
impossibility proof to resolve these points in its favor with
convincing arguments, which have {\em not} been provided. Indeed, all
three protocols QBC1 to QBC3, and even protocol QBC01 to a lesser
extent, lie outside the framework of the impossibility proof, and no
impossibility argument has been given for this kind of protocols. While there
are various underlying reasons on the limited scope of the impossibility proof formulation, a
major one is that the interactive flow of information between Adam and
Babe may prevent cheating because of each party's lack of relevant
information at any particular stage of the protocol. Such information
flow is what makes the Yao model of two-party protocols \cite{yao} not
sufficiently specific to characterize all QBC protocols, which he did
not claim to have done.  Furthermore, modification of the Yao model to
have measurements at the end of the commitment phase, perhaps thought
to be equivalent by the Lo-Popescu theorem \cite{lo_pop}, is not
justified with the use of anonymous states because the state needs to
be known to guarantee the validity of that theorem.  The basic problem
of a general impossibility proof lies in the characterization of the
essence of any possible QBC protocol that makes it insecure.  The
information flow problem that makes it so difficult to characterize
all classical protocols surely carries over to the quantum domain.

There are well-known and widely accepted claims in the literature
\cite{crepeau}-\cite{crepeau1} that classical noisy channels would
make unconditionally secure bit commitment possible.  While
I believe the specific protocols described in
\cite{crepeau}-\cite{crepeau1} are not proved unconditionally
secure, I also believe unconditionally secure ones can indeed be based
on noisy channels, a subject to be discussed elsewhere.  Such results are {\em not} considered to be {\em contradictory} to the
QBC impossibility proof presumably for the following reasons.  First, classical
noise is often thought to be part of an imperfect channel, i.e. it
does not have to be present in principle.  Apart from the points made
at the beginning of this section, such a viewpoint is not correct.  The quantum noise in any given quantum signaling scheme
for classical communication, the minimum amount of which is determined
through the optimum quantum measurement via quantum detection theory,
is {\em in principle} unavoidable and functions exactly like classical
noise in the optimal quantum detector
\cite{ykl}-\cite{yuen_squeezed}. As will be shown elsewhere, this
crucial point opens up the possibility of developing unconditionally
secure, practical, and efficient
optical-speed cryptographic systems for all the standard cryptographic
objectives via quantum states that are not superpositions of one another.  Secondly, a truly classical noise system would not entail
the possibility of quantum entanglement and EPR cheating.  However,
there are many ways to suppress EPR cheats, such as the example in
Section I and the QBC2 in Section VI.  While it is not easy to restore
unconditional security with such suppression in a perfect channel,
a noisy channel, even one created with quantum noise, would provide a powerful
way for such restoration.  Indeed, the development of such protocols
will be the subject of a future treatment.

\newpage
\section{\hspace{.2in}Protocol QBC1}

In this Section we consider the use of anonymous states in a QBC
protocol which is essentially the one in version 1 (v1-v3) of this
paper.  In this protocol QBC1, the bit value is encoded in the parity
of a sequence as in QBC0 of Section II, except that each individual
state is obtained with Adam applying the openly known $U_0$ or $U_1$
to the states $|\psi\rangle$ sent to him by Babe, corresponding to the
0 or 1 bit position in the sequence.  For example, $| \psi \rangle$
could be any state on a fixed great circle of the Bloch-Poincar\'e
sphere of a qubit, with $U_0 = I$ and $U_1$ being a rotation by a
fixed angle on the great circle independently of the bit position,
say with $\langle \psi | U^\dag_1 U_0 | \psi \rangle = \lambda > 0$.
See Ref.~\cite{yuen_capri} for further discussion of anonymous-key
cryptography.  Coherent-state implementation is also possible, as in
QBC01.

{\em PROTOCOL} QBC1:

(i) Babe sends Adam a sequence of $n$ qubit states $|\psi_l\rangle \in
{\mathcal H}^B_l$, ${\mathcal H}^B = \bigotimes_l {\mathcal H}^B_l$,
$l \in \{1,\ldots,n\}$, unknown to Adam.

(ii) Adam commits via the parity of the sequence ${\bf
j}=(j_1,\ldots,j_n) \in \{0,1\}^n$ by applying $U_{lj_l}$ to
$|\psi_l\rangle$ for openly known $U_{l0}$ and $U_{l1}$, with $\langle
\psi_l | U^\dag_{l1}U_{l0}|\psi_l\rangle = \lambda > 0$ independently
of $l$.

(iii) Adam opens by revealing his ${\bf j}$ sequence.  Babe checks
every state $U_{lj_l}|\psi_l\rangle$.

This scheme can be made concealing exactly as in QBC0, (14)-(15).  As for its
binding behavior, consider first the situation in which Adam can only
entangle each qubit individually. He cannot switch any committed
$U_{l0}|\psi_l\rangle$ or $U_{l1}|\psi_l\rangle$ to any other state
due to local state invariance which applies to each of the states he
sends separately for {\em that} state, expressing the obvious fact
that there is no entanglement to a single state.  If he were to
entangle $U_{l0}|\psi_l\rangle$ or $U_{l1}|\psi_l\rangle$ to another
state anyway, he would just present a mixed state for that qubit to
Babe for that $j_l$.  In this case, a different criterion needs to be
used as discussed below.  If he sticks to committing first a correct
state for the bit, the best cheating probability he can get it
\begin{equation}
P^A_c = \left| \langle \psi_l | U^\dag_{l1}U_{l0}|\psi_l\rangle
\right|^2 = \lambda^2
\end{equation}
by generating any sequence of $n-1$ states, picking the last one for
the bit commitment, and declaring it to be otherwise when desired.
From (15) and (23), one can make $P^A_c = O(m^{-1})$ and $\bar{P}^B_c -1/2 =
O(2^{-m})$ with $n = O(m^2)$.  Hence unconditional security is
obtained for large $m$ if (23) is indeed the overall best Adam can
do. In addition to $P^A_c$, one can use another criterion, $P^A_a$,
the average probability that Adam's committed evidence is accepted by
Babe after he opens, which is always at least 1/2 similar to $P^B_c$
with $P^A_a=(1+P^A_c)/2$ when (1) is used as an initial state
$|\bar{\Phi}_0 \rangle$ by Adam.  For a general $| \bar{\Phi}_0
\rangle$, (6) can be simply generalized to give an expression for
$P^A_a$ with optimization for $\bar{P}^A_a$ to be performed also over
initial $\{\phi^0_i\rangle\},\{p^0_i\}$.  In the present situation,
since single-qubit entanglement by Adam would just lead to a mixed
presented state from local state invariance, $\bar{P}^A_a$ is obtained
by a fixed $|\phi^0_i\rangle=|\phi^0\rangle$ with $\bar{P}^A_a =
(1+\lambda)/2$.  Thus, $\bar{P}^B_c-\frac{1}{2}=O(2^{-m})$ and
$\bar{P}^A_a-\frac{1}{2}=O(m^{-1})$ are achieved for $n=O(m^3)$.

Adam can, however, form the entanglement without knowing the
$|\psi_l\rangle$'s, by applying the unitary operator $U$ on ${\mathcal
H}^A \otimes {\mathcal H}^B$,
\begin{equation}
U=\sum_i |e_i\rangle\langle e_i| \otimes U_i
\end{equation}
with initial state $|A\rangle \in {\mathcal H}^A$ satisfying
$\sqrt{p_i}=\langle e_i | A \rangle$, as was indicated in version 1 of
this paper.  On the other hand, contrary to the claim in that version,
Adam can also entangle qubit by qubit via, for each $i =
(i_1,\ldots,i_n)$ in (25),
\begin{equation}
U_i = \bigotimes_l U_{li_l} = (I_1 \otimes \ldots \otimes U_{ni_n})
\ldots (U_{ii_1} \otimes \ldots \otimes I_n).
\end{equation}
By applying (24)-(25), Adam can form the proper entangled state (1) or
(2) without knowing the $|\psi_l\rangle$'s.  However, he cannot
determine the cheating transformation $U^A$.  In general such a
cheating transformation for the $\rho^B_0 \neq \rho^B_1$ case is
determined by Uhlmann's theorem as follows \cite{jozsa}.

Let $|\lambda_i\rangle$ and $|\mu_i\rangle$ be the eigenstates of
$\rho^B_0$ and $\rho^B_1$ with eigenvalues $\lambda_i$ and $\mu_i$.
The Schmidt normal forms of the purifications $|\Phi_0\rangle$ and $|
\Phi_1 \rangle$ of $\rho^B_0$ and $\rho^B_1$ are given by
\begin{equation}
|\Phi_0\rangle = \sum_i \sqrt{\lambda_i} |f_i\rangle |\lambda_i\rangle,
\end{equation}
\begin{equation}
|\Phi_1\rangle = \sum_i \sqrt{\mu_i} |g_i\rangle |\mu_i\rangle
\end{equation}
for complete orthonormal sets $\{|f_i\rangle\}$ and $\{|g_i\rangle\}$
on ${\mathcal H}^A$.  Define the unitary operators $U_0$, $U_1$ and
$U_2$ by
\begin{equation}
U_0|\lambda_i\rangle = |\mu_i\rangle,
\end{equation}
\begin{equation}
U_1|\lambda_i\rangle = |f_i \rangle,
\end{equation}
\begin{equation}
U_2|\mu_i\rangle=|g_i \rangle.
\end{equation}
Since one can always pick ${\mathcal H}^A$ to be isomorphic to
${\mathcal H}^B$, one can identify them via the isomorphism.  Let
$U$ be the unitary operator for the polar decomposition of
$\sqrt{\rho^B_0} \sqrt{\rho^B_1}$ \cite{note_infdim},
\begin{equation}
\sqrt{\rho^B_0} \sqrt{\rho^B_1} = \left| \sqrt{\rho^B_0}
\sqrt{\rho^B_1} \right| U.
\end{equation}
Then $\left| \langle \Phi_0 | \Phi_1 \rangle \right|^2$ assumes its
maximum value $F(\rho^B_0,\rho^B_1)$ when
\begin{equation}
U U^T_2 U_0 U_0^T U^T_1 = I
\end{equation}
where $T$ denotes the transpose operation. Thus, when $\rho^B_0$, $\rho^B_1$, and $|e_i\rangle$ are given,
$|g_i \rangle = |e'_i\rangle$ of $|\Phi_1\rangle$ is determined from (30) via solving for $U$
from (32), which required detailed explicit knowledge of $\rho^B_0$
and $\rho^B_1$. In terms of the notation for (13)-(14), the density
operators are
\begin{equation}
\rho^B_i = \frac{1}{2^{n-1}} \sum_{{\bf j} \in \Lambda_i} \,
\bigotimes^n_{l=1} U_{lj_l} |\psi_l\rangle \langle \psi_l |
U^\dag_{lj_l}\ \qquad i \in \{0,1\},
\end{equation}
which is unknown to Adam through the $|\psi_l\rangle$ uncertainty. If Adam picks
a cheating transformation for a particular $|\psi_l\rangle$ sequence,
and then the $|\psi_l\rangle$ sequence is randomly varied, it is
easily seen that the resulting $P^A_c$ can be very small, as e.g. when
the corresponding odd-parity state is actually of even parity.
However, it is not easy to develop an unconditional security proof because Adam has many
other possible actions, including committing states which are not
exactly correct for the bit value as mentioned above. Nevertheless, the
protocol clearly shows in a simple way that the impossibility proof
fails to work as intended.  Note that this anonymous-key strategy also
works in the case $\rho^B_0 = \rho^B_1$ if $\rho^B_0$ is not highly
degenerate, e.g., not proportional to the identity $I^B$, such that its
eigenstates cannot be readily determined as in the case of (33).  Indeed, for
$n=\infty$ the $\rho^B_0$ and $\rho^B_1$ from (33) are equal and not
proportional to $I^B$. Note that the strategy of this protocol, namely
the use of anonymous states, is applicable to {\em any} QBC protocol,
and will be employed next for protocol QBC2.

\newpage

\section{\hspace{.2in}Protocol QBC2}

In this section, protocol QBC2 is developed with a complete
unconditional security proof by exploting the following point: the
states $\rho^B_0$ and $\rho^B_1$ that enter into (7) are not
necessarily the marginal states obtained from (1)-(2) due to Babe's
lack of information built into a QBC protocol.  This situation is
actually  easy to obtain, but then Adam can usually cheat successfully
with this information.  The anonymous-key technique can be utilized to
prevent both Adam and Babe from cheating to yield an unconditionally
secure  protocol to be
explained in successive steps as follows.

In anonymous-key encryption \cite{yuen_capri}, Babe transmits to Adam
a state $| \psi \rangle$ only known to herself.  Adam sends a bit b
back to Babe via modulating $| \psi \rangle$ by openly known unitary
operators $U_b$.  For the present purpose, the following would suffice
-- $| \psi \rangle \in S_0$ is one of the four possible BB84 states of a qubit,
$S_0 = \{|\uparrow\rangle$, $|\rightarrow\rangle$, $|\nearrow\rangle$,
$|\searrow\rangle \}$, (e.g. the vertically, horitzontally, and
diagonally polarized states).  Adam sends back $U_b|\psi \rangle$
with $U_0=I^B$ and $U_1$ being a clockwise rotation by $\pi/2$ on the
polarization circle, so that Babe can always tell the bit from the
state.  Let her send Adam a set $S$ of the above four different states
on four qubits in a random order known only to herself, with each
state {\em named} by its order.  Thus, $S=\{ |\lambda_1\rangle_1,
|\lambda_2\rangle_2, |\lambda_3\rangle_3, |\lambda_4\rangle_4 \}$
where the subscript $j$ on $|\,\,\rangle_j$ denotes the name of the state
and $\{\lambda_j\}$ is a random permutation of the set ${\mathcal S}_0=\{\uparrow,
\rightarrow,\nearrow,\searrow\}$.  Adam picks randomly one of these
four named states in $S$, keeping the name to himself, modulates it and sends
it to Babe as the commitment.  For example, he chooses
$|\lambda_2\rangle_2$ with subscript 2 on $|\,\,\rangle_2$ known to him,
rotates $\lambda_2$ unknown to him clockwise by $\pi/2$ for b = 1,
and sends it back to Babe who does not know the state name "2" yet.
He opens by revealing the state name and the bit value. Without
knowing the state name, it is easy to check that $\rho^B_0 = \rho^B_1
= I^B/2$ for Babe.  When she learns the state name from Adam's
opening, she knows the corresponding state for each bit value and can
verify by measuring the corresponding projection. The actual
permutation of the $S_0$-states in $S$ has to be hidden from Adam
because if he knows, he can cheat by committing any state in $S$ and
announcing it to be another appropriate state from $S$.

Consider first Adam's possibility of cheating.  When he picks a
specific named state $|\lambda_j\rangle_j$, he cannot apply the EPR
cheat as a consequence of local state invariance or the fact that
there is no entanglement for a singla state.  He can announce a
different name of the state from the one he actually sent, with a
probability of successfully reversing b (i.e. getting it accepted by
Babe in her verification) given by 3/4..  He can use his own state
instead of the one sent by Babe; the best way to do that is by trying to determine which name
corresponds to which state in $S$ by optimally processing the set $S$
from $M$-ary quantum detection theory (cf. Appendix A).  In each case
he attains a probability of success bounded away from zero.  Let $p_A$
be his maximum probability of success, which is determined by the
optimal $M$-ary quantum detector because his openings amount to a
decision making that consists in matching each $\lambda_j$ with an
element of ${\mathcal S}_0$.  The exact value of $p_A$ is not relevant for the
security proof of our final protocol.  The only relevant point here is
that $p_A$ is a fixed number less than one.  Hence, in an
independent $m$-sequence, his probability of successful cheating,
$\bar{P}^A_c = p^m_A$, goes to zero exponentially in $m$.

To show that $p_A <1$, assume that Adam can cheat perfectly with $p_A
= 1$. This implies that he can determine $\lambda_j$
for each $|\lambda_j\rangle_j$ from the set $S$ with certainty without knowing the
random permutation.  However, the different possible permutations
yield nonorthogonal (mixed) states on the different qubit sets.  By
Theorem A2 in Appendix A, $p_A = 1$ is impossible.  Indeed, the
optimum $p_A$ is a fixed number bounded away from zero, not
arbitrarily small in a parameter $n$ that grows with the number of such
randomly permuted four-state sets.

It is possible for Adam to consider EPR cheats by permuting the
contents of the states to be used later with a single qubit while keeping track of the
state name.  In this way, he can form the entanglement (1)-(2), but he
cannot transform one into the other without knowing the specific
permutation of the $\lambda_j$ in the set $S$ presented to him.  And,
of course, if he knows the permutation, he can cheat by proper
announcement without the need for entanglement.  Note also that if he
can entangle and transform without knowing the actual permutation,
local state invariance would be violated by permuting the states back
to the given order.  Indeed, this and {\em all} other possible attacks
by Adam are accounted for in the above argument that $p_A < 1$ holds
in any of Adam's possible cheating schemes as a consequence of optimum
quantum detection theory.

The only way that Babe can cheat is to send Adam a different set $S'$
of states, e.g. the same polarization state on the polarization circle
for all four qubits, which would yield $\bar{P}^B_c = 1$.  This is to
be prevented statistically via testing by having Babe send Adam a
total number of $n$ sets of $S$-states, all named by their order.
Consider first the case in which Adam only commits a single qubit, and
Babe sends a total of $4n$ states $|\lambda_{jl}\rangle$, $j \in
\{1,\ldots,n\}$, $l \in \{1,\ldots,4\}$.  If Babe is honest, then, for
each $j$, $\{ |\lambda_{jl}\rangle\}$ is a random permutation of $S_0$.  To prevent
Babe from cheating, Adam would randomly set aside one set $j_0$ and
ask Babe for the state identities in the other $n-1$ sets.  After Babe
reveals the state identities from their names provided by Adam, he can
verify that Babe indeed sent him sets of proper states and proceed to
pick one from the $j_0$ set to commit his bit.  If Babe sends a set
$S' = \{ |1\rangle, |2\rangle, |3\rangle, |4\rangle \}$, which is not
a random permutation of $S_0$, then there is a probability $p_1$ that it will pass Adam's testing
verification,
\begin{equation}
p_1 = | \langle 1 |~\uparrow \rangle |^2 \cdot | \langle 2|\rightarrow \rangle |^2 \cdot | \langle 3 |\nearrow \rangle |^2 \cdot | \langle
4 |\searrow \rangle |^2,
\end{equation}
and a corresponding optimum probability $p_2$ that Babe can determine
the bit knowing the qubit is from $S'$. The value of $p_2$ is
determined by the optimum binary quantum detector. For example, if Babe sends
all states at the angle $\pi/8$ from $\uparrow$, $p_1=\cos^2
\frac{\pi}{8} \cos^2 \frac{\pi}{8} \cos^2 \frac{3\pi}{8} \sin^2
\frac{\pi}{8}$ and $p_2=1$.  As far as the existence of an
unconditionally secure protocol is concerned, the only thing we need
to know is that $p_1=1$ implies $p_2=\frac{1}{2}$ from (34).  It is clear that any entanglement used by Babe on
the state she sent would not help her cheat, because Adam is doing
everything on the individual qubit level determined by the individual
marginal qubit states. Indeed, Babe's entanglement would only make
$p_2$ smaller. For Babe to get $\bar{P}^B_c$ away from
$\frac{1}{2}$, she needs to send state sets with $p_2 =
\frac{1}{2}+\epsilon$ where $\epsilon$ is bounded away from zero,
i.e. not arbitrarily small as a function of $n$, and send enough
of them so that the chance that one of them is picked as
$j_0$ by Adam is also not arbitrarily small for large $n$.  In such a
situation where Adam retains one of Babe's cheating state sets which
constitute a nonzero fraction $\gamma$ of the total number $n$, the
probability that Babe's cheating would not be found out is $p^{\gamma
n-1}_1$, assuming Adam indeed sets aside one of the cheating state
sets, which goes to zero exponentially.  This argument is essentially
correct and will be presented rigorously in the more general situation of the
protocol in the following.  Here we tried to indicate the simple
intuitive picture of the situation, and the fact that our scheme so
far already contradicts the (IP) statement, although it falls short of
the (US) statement.  It should be evident that regardless of whether
(US) can be obtained in this kind of schemes, the are {\em not}
covered by the formulation of the impossibility proof.

Were Babe found to be cheating, the protocol would of course abort,
which is equivalent to one party aborting in the middle of any
protocol, something each party can always choose to do.  Thus, our
scheme is no different in this respect from any other cryptographic
protocol and is essentially different from the cheat-sensitive QBC protocols
\cite{hardy} in that it has nothing to do with detecting possible
cheating by Adam and Babe after Adam commits as prescribed in the
definition of cheat-sensitive protocols. Indeed, Adam can discover the cheating before he commits
the bit. Even though he could postpone the cheating detection
measurement in our protocol, such a move would have betrayed his bit
to Babe, cf. point (4) in Section IV. More significantly, the cheating probabilities were not
quantified precisely in Ref.~\cite{hardy} -- presumably if the successful
cheating probability is bounded away from zero, then so is the
cheat-detection probability. In the present case, arbitrarily small
successful cheating probabilities can be obtained in the next
protocol, the parameters $n,m$ of which are determined as shown in the
following security proof.

{\em PROTOCOL} QBC2:

(i) Babe sends Adam $n$ sets of qubit states, each set a random
permutation of the four BB84 states on four different qubits,
in a random order only known to herself.  The states are named by
their order in the sequence.

(ii) Adam randomly puts $m$ sets of such states aside and asks Babe to
identify the rest of the states from their names.  After checking that
the states are correct, he commits the bit by picking one state
randomly out of each of the $m$ sets, modulates them by the same
$U_b$, and sends them to Babe.

(iii) Adam opens by revealing the names of the states he sent and the
bit value.  Babe verifies by measuring the corresponding projections.

It should be clear that no entanglement cheating would be effective in this
protocol: as discussed above, entanglement cheating by Adam or Babe
serves no purpose as the qubits are processed individually. For each
bit value Adam commits, there is only one product marginal state for
Babe and thus no cheating transformation for Adam.  If Adam entangles
anyway, he would merely send back mixed marginal states to Babe as she
verifies on individual qubits. If he does not commit a correct state
as discussed after (23) in Section V, it merely changes $p_A$, the
optimum value of which is not 1 as shown above.  If Babe entangles
anyway, she would just get back mixed states for herself.  Consequence
of such a situation, however, is also covered in the following. Similarly,
introducing any classical correlation would serve no purpose. The
protocol is binding because Adam's $\bar{P}^A_c = p^m_A \rightarrow 0$
for large $m$.  It is concealing basically for the same reason as the
single-qubit case, a systematic proof given as follows.

Let $N$ be the number of state sets Babe sends to Adam with
probabilities $p_1$ of passing Adam's detection, $p_1 < 1$ with
corresponding $p_2 > \frac{1}{2}$.  Consider first the case in which these probabilities are uniform
among the $N$ sets so that Babe can have the best possible $p_2$ given
$p_1$ among the $m$ different committed qubits. 
The other $n-N$ sets have $p_1=1$ and $p_2 = \frac{1}{2}$.  The probability that $k$ of these
$N$ sets fall into the $m$ choices by Adam is given by the
hypergeometric distribution,
\begin{equation}
P_k(N,n,m) = 
\frac{\left(
\begin{array}{c}
N \\ k\end{array}\right) \left(
\begin{array}{c}
n-N \\ m-k\end{array} \right)}{\left(
\begin{array}{c}
n \\ m\end{array} \right)}.
\end{equation}
The probability that none of these $N$ sets fall into the chosen $m$
group is $P_0(N,n,m)$, a decreasing function of $N$ and an increasing
function of $n$.  Let $m$ be the smallest integer that yields $\bar{P}^A_c = p^m_A \le
\epsilon$ for given $\epsilon > 0$.  The idea is that $N$ must be
large enough that at least one of the $N$ sets needs to fall into the
$m$ group to get $\bar{P}^B_c > \frac{1}{2}$, but then by making $n$
large, $N$ would have to be so large that the probability $P_u$ that
Babe's cheating sets are undetected becomes too small. Recall that
$\bar{P}^B_c$ is the optimal probability Babe succeeds in identifying
the bit from measurements on $m$ committed qubits. It will be shown
that the condition
\begin{equation}
\bar{P}^B_c \ge \frac{1}{2} + \epsilon
\label{cond}
\end{equation}
would imply $P_u \le \epsilon$ by proper choice of $n$, thus ensuring
unconditional security. Since Babe must have at least one of the $N$ sets picked up by Adam
among his $m$ sets in order to satisfy (\ref{cond}),
\begin{equation}
\bar{P}^B_c \le \frac{1}{2} P_0 + (1-P_0) P(p_2,m) \le
1-\frac{P_0}{2}.
\label{upbound}
\end{equation}
By equating the upper and lower bounds (\ref{cond}) and
(\ref{upbound}) on $\bar{P}^B_c$, $N$ must satisfy
\begin{equation}
N \ge f(\epsilon,n,m(\epsilon,p_A))
\label{Nbound}
\end{equation}
where $f$ is defined through $P_0(N,n,m)$ and is an increasing
function of $n$.  For any $N,n$, 
\begin{displaymath}
P_0(N,n,m)=
\frac{\left(
\begin{array}{c}
n-m \\
N
\end{array} \right)}{ \left(
\begin{array}{c}
n \\
N
\end{array} \right)}
\end{displaymath}
can be made arbitrarily small with $n$
large.  Thus, $N$ can be forced to be arbitrarily small from
(\ref{Nbound}) with $n$ sufficiently large. If there is an a priori
maximum $\bar{p}_1$ among the qubits in the $N$ sets, which is proved
in the following, one would have $P_u \le
\bar{p}^{N-m}_1$. So $n$ can be chosen to make $N$ large enough from
(\ref{Nbound}) to yield $\bar{p}^{N-m}_1 = \epsilon$.  As a
consequence, $\bar{P}^A_c \le \epsilon$ and $\bar{P}^B_c \le
\epsilon$, proving (US).

To put an a priori limit on $p_1$ independent of $n$ and less than
one, consider first the case where all qubits in the $N$ sets have the
same underlying $S'$ so that Babe knows what measurement to make on
each. Let $P(S',m)$ be the optimum probability that Babe succeeds in identifying b from measurements on the $m$ qubit
sets. Thus, $P(S',m)$ is a continuous function of the $S'$ that gives
rise to the $p_2$ as it is a trace norm of the states from (A4). (All norm topologies are equivalent in finite-dimensional spaces). In order
for (\ref{cond}) to be satisfied, one must have
\begin{equation}
P(S',m) \ge \frac{1}{2}+\epsilon
\label{cond2}
\end{equation}
for some $\epsilon > 0$.   The maximum $p_1$ that Babe can have is determined
among all the qubit sets $S'$ that satisfy
(\ref{cond2}) and $1 \ge P(S',m)$. The maximum $\bar{p}_1 =
\max_{S''} p_1(S'')$ exists for the following reason.
Thus the set of $S'$ obeying (\ref{cond2}) and $P(S',m) \le 1$ is closed and thus compact.  The function $p_1(S')$ of
(34) is continuous.  The existence of $\bar{p}_1$ thus follows from
the Weierstrass theorem. That is, a maximum $\bar{p}_1$ is achieved by
some $S'_0$ in the constraint set and so $\bar{p}_1 < 1$. Now suppose
Babe has formed entanglements among the sets she sends to Adam.  The
$N$ sets are defined according to whether each marginal state, as checked and
modulated by Adam, would have $p_{j1}=1,$ $j \in \{1,\ldots,n\}$.
Thus, instead of $P(S',m)$ one has $P({\mathbf S}',m)$ that includes
optimization over all possible entangled states ${\mathbf S}'$, which
provides an upper
bound to $\bar{P}^B_c$ and is still given through the trace norm
(A4). Let $\bar{p}_1 = \max_{S_j} p_1(S'_j)$ for all marginal $S'_j$ obtained
from ${\mathbf S}'$ that satisfy $1 \ge P({\mathbf S}',m) \ge
\frac{1}{2}+\epsilon$.  All $S'_j$ in the $N$-set lead to $p_{j2} <
\frac{1}{2}$ by definition of the $N$-set.  Thus, the existence of
$\bar{p}_1 < 1$ follows as in the uniform $S'_j$ case.  We have now exhausted all possible actions
by Adam and Babe.

In order to execute this protocol in accordance with the above proof
in choosing $m$ and $n$, one needs to know $p_A$ and $\bar{p}_1$.  These appear to be difficult to
obtain analytically, and numerical solutions would need to be used in
an actual implementation. In such a situation, the above technicality
on the existence of $\bar{p}_1 < 1$ would not occur.  While it is easily shown that no four
large-energy coherent states can approximate the behavior of the four
BB84 states in $S_0$, it may still be possible to develop large-energy
coherent-state implementation of this protocol because not all
properties of the BB84 states are needed.

\newpage

\section{\hspace{0.2in}Protocol QBC3}

The points (7)-(8) in Section IV are now exploited to create a
protocol that defeats Adam's EPR cheat.  Consider the following
addition to protocol QBC0 in Section II: after Adam commits, Babe
picks randomly $N$ out of the $n$ qubits and measures randomly on each
either $| \phi \rangle \langle \phi |$ or $| \phi' \rangle \langle
\phi' |$, but does not tell Adam which qubits she picked and what
measurement results she obtained.  When Adam opens, she would verify
among the $N$ qubits those that match Adam's announcement and the rest
$n-N$ qubits, and take those that don't match Adam's announcement as
correct.  Thus, she does not have a perfect verification, but Adam
cannot cheat successfully by changing one bit position in his
announcement when $N/n$ is small. On the other hand, this action by
Babe effectively destroys the entanglement that Adam may have formed
for the EPR cheat, as shown below.  Babe needs to keep secret which
$N$ qubits she made measurements upon, or else Adam can alter his
basis $|e_i \rangle$ to entangle properly to the other $n-N$
qubits. Condition on the parameters will be given.

{\em PROTOCOL QBC3}

(i) Adam sends Babe a sequence of $n$ qubits, each in either one of
$\{ |\phi\rangle, |\phi'\rangle \}$, and commits b via the parity of
the sequence with uniform probability.

(ii) Babe randomly picks $N$ out of $n$ qubits, randomly measures
either $| \phi \rangle \langle \phi |$ or $| \phi' \rangle \langle
\phi' |$ on each, and keeps the results secret from Adam.

(iii) After Adam reveals the sequence commitment, Babe verifies those
states that match among the $N$ measured qubits and the $n-N$
unmeasured ones.

The protocol can be made concealing as in QBC0 and QBC1, but Adam can now cheat
in more ways.  Similar to QBC1, he can pick one qubit and announce it
otherwise, which now has a higher probability of success because of
Babe's measurements. From the union bound on the probability of two
possible events,
\begin{equation}
P^A_c \le | \langle \phi | \phi' \rangle |^2 + \frac{N}{n}.
\label{pac_upbound}
\end{equation}
Thus one may pick
\begin{equation}
| \langle \phi | \phi' \rangle |^2 = O(m^{-1})
\label{pick_overlap}
\end{equation}
similar to (23) and also
\begin{equation}
N/n = O(m^{-1}),
\label{qubitfrac}
\end{equation}
so that $P^A_c = O(m^{-1})$.  Adam can, in view of Babe's possible
measurements, entangle as small a number of qubits as possible.  If he
wants an entanglement cheating probability of
\begin{equation}
\bar{P}^A_c = 1 - O(m^{-1}),
\label{cheat_prob}
\end{equation}
he would need to entangle $n' = O(m \log m)$, so that the resulting
$\bar{P}^B_c = \frac{1}{2}+O(m^{-1})$ would guarantee
(\ref{cheat_prob}) through (12).  Thus, to maintain just this order of
$\bar{P}^B_c$, $n$ should be reduced to $n=O(m \log m)$ compared to
QBC1, and so $N = O(\log m)$ from (43).

If Adam just cheats as if Babe has made no measurement, a
direct computation shows that, for $p_i = p'_i = 1/M$,
\begin{equation}
| \langle \Phi''_0 | \Phi_1 \rangle |^2 = 2^{-N}
\label{entoverlap}
\end{equation}
where $|\Phi''_0\rangle$ is the cheating entangled state from
$|\Phi_0\rangle$ after Babe made her $N$ qubit measurements as
follows.  The state $|\Phi''_0\rangle$ can be written, from (5),
\begin{equation}
|\Phi''_0\rangle = \sqrt{\frac{\mathcal N}{M}} \sum_j \sum_{i_{N+1}\ldots
i_{n-1}} V_{{\bf i}_N i_{N+1} \ldots i_{n-1}j} |e'_j \rangle
|\phi_{{\bf i}_N i_{N+1} \ldots i_{n-1}} \rangle
\end{equation}
where ${\bf i}_N$ are the fixed indices corresponding to Babe's
measurements results and ${\mathcal N}$ is a normalization constant determined to
be ${\mathcal N} = 2^N$. Then (44) follows from (45) and the unitarity
of $V_{ij}$.  However, this does not yet constitute an unconditional security
proof for the following reasons.  Adam does not have to
apply the cheating transformation as if Babe has made no
measurements. It remains to be demonstrated that his optimal cheating
transformation, particularly in the case he does not generate an
exactly correct initial state for the bit value as discussed after
Eq.~(23), would lead to an arbitrarily small $P^A_c$.
Furthermore, Adam may aim lower than $\bar{P}^A_c = 1 - O(m^{-1})$ by
optimizing differently, just to defeat (US).

I believe QBC3 is in fact unconditionally secure, as I believe QBC1
is, and a new formulation of the QBC problem is being developed to
facilitate further analysis of the $P^A_c$ behavior in QBC protocols
with possible entanglement attacks.  Such general treatment is important
because the strategy of this protocol is applicable to {\em all} QBC
protocols in which the bit value is obtained from a correlated
function of the individual bit positions, and the strategy of QBC1,
namely the use of anonymous states, is applicable to {\em any} QBC protocol.

\newpage

\section{\hspace{0.2in}Conclusion}

I hope this paper leaves no doubt that not only is there no general
impossibility proof for unconditionally secure quantum bit
commitment, but that, in fact, an unconditionally secure QBC protocol
has actually been provided.  The intuitive reasons and a complete
proof that QBC2 satisfies (US) have been described in Section VI.  The
protocols QBC1 and QBC3, while not proved to be unconditionally secure
in this paper, already demonstrate the failure of the impossibility
proof given in the literature.  Additional gaps of the impossibility
proof are indicated in Section IV and can be exploited for further
secure QBC schemes.

Some comments on the practicality of our protocols are in order.
Protocols QBC0, QBC1, and QBC3 can be readily implemented with
large-energy coherent states.  However, there is a sensitivity problem
that results from $|\langle \phi|\phi' \rangle| \sim 0$, which
obscures the difference in practice between the two cases of detection
for verification versus cheating corresponding to the cases when the
state is known or unknown.  An investigation into sensitive detection
schemes would be timely.  Also, it is expected that this and other
practical difficulties can be alleviated by the use of
error-correcting codes or hash functions more complicated than
parity.  Perhaps a large-energy coherent-state scheme similar to QBC2
can also be developed.  Another promising avenue is the utilization of
the irreducible quantum noise in quantum signal detection schemes to achieve
unconditionally secure bit commitment.  The loss in fiber-optic
communications, especially for the established Internet backbone, can
also be used to generate irreducible quantum noise.  The resulting
protocols, together with similarly possible quantum key distribution
and encryption schemes, may open the exciting possibility of
optical-speed unconditionally secure cryptography for widespread applications.

\newpage

\addcontentsline{toc}{section}{Acknowledgment}

\section*{Acknowledgment}

I am indebted to many colleagues for their criticism, support, and
discussions, especially to M. d'Ariano, C. Bennett, H. Lo, D. Mayers,
T.Mor, and M. Ozawa, but also to G. Barbosa, S. Barnett, G. Brassard,
H. Bernstein, J. Bub, H. Chau, G. Gilbert, N. Gisin, P. Grangier,
O. Hirota, R. Jozsa, B. Leslau, P. Kumar, S. Kumar, S. Lloyd, C. Macchiavello, L. Maccone, M. Rubin,
J. Shapiro, Y. Shih, and P. Tombesi.

This work was supported in part by the Defense Advanced Research
Project Agency and in part by the Army Research Office.

\newpage

\addcontentsline{toc}{section}{Appendix A: Quantum Detection Theory}

\renewcommand{\theequation}{A\arabic{equation}}
\setcounter{equation}{0}

\section*{Appendix A \\
Quantum Detection Theory}

Quantum detection theory \cite{helstrom}, \cite{ykl} is concerned with
the determination of the optimum quantum measurement and the resulting
optimum performance for discriminating a finite number $M$ of
alternative hypotheses according to a given performance criterion
linear in the density operators $\rho_j$, $j \in \{1,\ldots,M\}$,
describing the quantum states of the different alternatives.  It has
not been used in the previous quantum cryptography literature other than my papers \cite{yuen_capri}, \cite{yuen_highrate}, \cite{yuen_crypt}, \cite{yuen_y2k},
although it actually has a crucial role, especially in QBC.  Thus,
Babe's optimum probability of cheating is given by the optimum binary
quantum detector for $\rho^B_0$ and $\rho^B_1$.

In binary quantum hypothesis testing with {\em a priori} probabilities
$p_0$ and $p_1 = 1 - p_0$, the decision is made on the basis of
measuring a POM (positive operator-valued measure) described by $\Pi_0$
and $\Pi_1 = I - \Pi_0$, $\Pi_0 \ge 0$ (an operator inequality $A \ge
B$ means that $A-B$ is positive semidefinite).  The hypothesis $i$ is
chosen correctly from the measurement result with probability ${\rm
tr} \Pi_i \rho_i$, so that the total probability of correct decison is
given by
\begin{equation}
P_{C2} = p_0 {\rm tr} \Pi_0 \rho_0 + p_1 {\rm tr} \Pi_1 \rho_1.
\end{equation}
In $M$-ary hypothesis testing, (A1) generalizes to
\begin{equation}
P_{CM} = \sum^M_{i=1} p_i {\rm tr} \Pi_i \rho_i
\end{equation}
where the $\{ \Pi_i \}$ form the $M$-outcome POM
\begin{equation}
\sum^M_{i=1} \Pi_i = I, \qquad \Pi_i \ge 0.
\end{equation}
An operator $\tau$ is called trace-class if its trace norm $\| \tau
\|_1 \equiv {\rm tr} \sqrt{\tau^\dag \tau}$ is defined (finite); thus
all operators on finite-dimensional spaces are trace-class. Density
operators are trace-class.  The
optimum $\bar{P}_{C2}$ among all POM's can be written as follows.  \\
\noindent{\em{Lemma A1}}:
\begin{equation}
\bar{P}_{C2} = \frac{1}{2} + \frac{1}{2} \| p_0 \rho_0 - p_1 \rho_1
\|_1.
\end{equation}
\\
\noindent{\em{Proof}}: Write $p_0 \rho_0 - p_1 \rho_1 = \sigma_+ -
\sigma_-$, the positive and negative eigenvalue parts, so that $|p_0
\rho_0 - p_1 \rho_1| \equiv \sqrt{(p_0 \rho_0 - p_1 \rho_1)^2} =
\sigma_+ + \sigma_-$. From ${\rm tr}(p_0 \rho_0 - p_1 \rho_1) = p_0 -
p_1$, one has ${\rm tr} \sigma_+ = {\rm tr}\sigma_- + p_0 - p_1$.
Now, from (A1),
\begin{displaymath}
\bar{P}_{C2} = p_1 + \max_{0 \le \Pi \le I} {\rm tr} \Pi (p_0 \rho_0 - p_1
\rho_1),
\end{displaymath}
while
\begin{displaymath}
\max_{0 \le \Pi \le I}{\rm tr}\Pi (p_0 \rho_0 - p_1 \rho_1) = \max_{0
\le \Pi \le I,\,\Pi \sigma_-=0} {\rm tr}\Pi \sigma_+ = {\rm tr} \sigma_+ = \frac{1}{2}
\| p_0 \rho_0 - p_1 \rho_1 \|_1 + \frac{1}{2} (p_0 - p_1),
\end{displaymath}
and (A4) follows. $\Box$
\\
\noindent{For two pure states, $|\psi_0\rangle$ and $|\psi_1\rangle$, (A4)
reduces to}
\begin{equation}
P_{C2} = \frac{1}{2} + \frac{1}{2} \sqrt{1 - 4 p_0 p_1 | \langle
\psi_0 | \psi_1 \rangle | ^2}.
\end{equation}

The use of ``information'' e.g. as in Ref.~\cite{bcjl}, is not
sufficient in QBC because it is not the relevant performance measure,
and the optimum detectors for $\bar{P}^B_c$ and mutual information are
usually not the same.  Indeed, generally in cryptography, the use of
mutual information is often not sufficiently precise because it has
only asymptotic significance in a noisy system, and at least Eve has
no possibility of coding.  Thus, the performance resulting from
attacks by Eve or by cheating among users in QBC should be measured by
their respective probabilities of success.  In some cases, including many quantum key-distribution situations, the mutual information
could be used to bound the successful eavesdropping probability.  But
even in those situations the resulting system design may be overly
pessimistic when the mutual information criterion is employed.

An important condition whose validity seems clear intuitively is that
$\bar{P}_{C2} =1$ in binary quantum detection if and only if the states
satisfy $\rho_0 \rho_1 = 0$, i.e. the ranges of $\rho_0$ and $\rho_1$
are orthogonal subspaces of the state Hilbert space.  The ``if'' part
is immediate and the ``only if'' part, which follows from (A5) when
$\rho_0$ and $\rho_1$ are pure states, seems to be a consequence of
the general no-clone theorem. Specifically, one would be able to clone
two nonorthogonal states if one could discriminate between them
perfectly.  However, the unitarity argument used for no-cloning is not sufficient to
include measurement transformations -- at least many physicists
believe that a quantum measurement transformation with a {\em
specific} reading is not describable by a unitary transformation on
any larger Hilbert space.  Nor is linearity sufficient.  Thus, the
impossibility of perfectly discriminating nonorthogonal pure states,
expressed as $\rho_0 \rho_1 \neq 0$ for general mixed states, is a
separate property to be demonstrated, indeed even just for {\em completing}
the no-clone argument.  That such a property can be demonstrated from
quantum detection theory, as done below, appears to me to be another
manifestation of the ``magical unity'' or consistency of the quantum
formalism.

The proof of the following theorem generalizes a finite-dimensional
proof for the case $\lambda_0 = \lambda_1=1$ first communicated to the
author by Masanao Ozawa.  \\
\noindent{\em{Theorem A1}}:  For positive constants
$\lambda_0,\lambda_1$ and density operators $\rho_0,\rho_1$, the
maximum value of $\| \lambda_0 \rho_0 - \lambda_1 \rho_1 \|_1$ amoung all possible
$\rho_0,\rho_1$ occurs only when $\rho_0 \rho_1 = 0$, with
\begin{equation}
\| \lambda_0 \rho_0 - \lambda_1 \rho_1 \|_1 = \lambda_0 + \lambda_1.
\end{equation}
\noindent{\em{Proof}}: In the finite-dimensional case, the polar
decomposition of
\begin{equation}
\rho' \equiv \lambda_0 \rho_0 - \lambda_1 \rho_1 = U | \lambda_0
\rho_0 - \lambda_1 \rho_1 |
\end{equation}
always exists for a unitary $U$.  In the infinite-dimensional case,
$U$ is only a partial isometry in general \cite{halmos}.  Since
$\rho'$ on ${\mathcal H}$ has an eigenvector decomposition as it is
trace-class, $U$ becomes an isometry when restricted to the space
${\mathcal H}_r \subset {\mathcal H}$, the range of $\rho'$.  Thus,
$U^\dag U = I_{{\mathcal H}_r}$, and we can write
\begin{equation}
|\lambda_0 \rho_0 - \lambda_1 \rho_1 | = V (\lambda_0 \rho_0 -
 \lambda_1 \rho_1)
\end{equation}
where $V = U^\dag$, $\| V \| = 1$.  From (A8),
\begin{equation}
\| \rho' \|_1 = {\rm tr} V(\lambda_0 \rho_0 - \lambda_1 \rho_1) = {\rm
tr} V\lambda_0\rho_0 - {\rm tr}V\lambda_1\rho_1.
\end{equation}
Now, from (\ref{lemma2}), for any trace-class operator $A \ge 0$ and any $V$ with
$\| V \| = 1$, the real part
\begin{displaymath}
{\rm Re\,tr}(VA) \le |{\rm tr}(VA)| \le \|VA\|_1 \le \| V \| \|A\|_1 =
{\rm Re\,tr}A,
\end{displaymath}
leading to
\begin{equation}
{\rm Re\,tr}(VA) \le {\rm Re\,tr}A
\end{equation}
from which it follows that
\begin{equation}
-\lambda_0 \le {\rm Re\,tr}V\lambda_0\rho_0 \le \lambda_0,\qquad - \lambda_1 \le
- {\rm tr} U\lambda_1 \rho_1 \le \lambda_1.
\end{equation}
From (A9) and (A11), we have
\begin{equation}
\max_{\rho_0,\rho_1}\|\rho'\|_1 = \lambda_0+\lambda_1
\end{equation}
which occurs when
\begin{equation}
{\rm Re\,tr} V\rho_0 = 1 \qquad {\rm and} \qquad {\rm Re\,tr} V\rho_1
= -1.
\end{equation}
Let $\rho_0$ have the spectral decomposition $\rho_0 = \sum_n \nu_n |
\phi_n \rangle \langle \phi_n |$.  Then (A13) implies
\begin{displaymath}
\sum_n \nu_n \langle \phi_n | V | \phi_n \rangle = 1.
\end{displaymath}
Since $0 \le \nu_n \le 1$ and $\sum_n \nu_n = 1$, if $\nu_n \neq 0$
we have ${\rm Re} \langle \phi_n | V | \phi_n \rangle = 1$ and hence
$V|\phi_n\rangle = |\phi_n\rangle$.  Let $\rho_1$ have the spectral
decomposition $\rho_1 = \sum_m \mu_m |\psi_m \rangle \langle \psi_m
|$. Similarly, if $\mu_m \neq 0$, then $V|\psi_m\rangle =
-|\psi_m\rangle$.  Since $VV^\dag = I_{{\mathcal H}_r}$, the
eigenvectors of $V$ with different eigenvalues are mutually orthogonal
and hence
\begin{equation}
\rho_0 \rho_1 = \sum_{m,n}\nu_n \mu_m |\phi_n \rangle \langle \phi_n
|\psi_m \rangle \langle \psi_m | = 0.
\end{equation}
$\Box$ \\
\noindent{{\em Corollary A1}}: $P_{C2} = 1$ if and only if $\rho_0\rho_1
= 0$.

I would like to emphasize that by itself, without the need for
unitarity, Corollary A1 already implies the no-clone theorem for
arbitrary $\rho_0 \rho_1 \neq 0$.  This is because if one can clone,
one can obtain an indefinitely large number of copies of the state,
which would make it possible to determine the state arbitrarily
accurately and hence contradicting the corollary.  On the other hand,
an argument using the physical interpretation of density operator as
an ensemble would show, in conjunction with the pure-state result from
(A5), that the eigenstates of $\rho_0$ and $\rho_1$ must be mutually
orthogonal to ensure $\bar{P}_{C2} = 0$, thus proving the corollary
{\em without} Theorem A1.  While this can be considered a {\em new}
kind of mathematics, proving mathematical theorems from physical
arguments, it is appropriate to separate physical interpretation from
what the mathematical formalism says by itself, if only to check
whether they are compatible.

The above theorem can be generalized to $M$-ary hypothesis testing.  \\
\noindent{{\em Theorem A2}}: $P_{CM}=1$ if and only if $\rho_i\rho_j =
0$ for all $i \neq j$.  \\
\noindent{{\em Proof}}: If one pair is not orthogonal, say $\rho_1
\rho_2 \neq 0$, then
\begin{displaymath}
P_{CM}=p_1 {\rm tr} \Pi_1 \rho_1 + p_2 {\rm tr} \Pi_2 \rho_2 +
\sum^M_{i=3}p_i {\rm tr}\Pi_i \rho_i \le p_1 {\rm tr}\Pi \rho_1 + p_2
{\rm tr}\Pi_2 \rho_2 + 1 - p_1 - p_2
\end{displaymath}
since (A3) implies $\| \Pi_i \| \le 1$ so that ${\rm tr}\Pi_i \rho_i \le 1$ from (18) and (A3). Because the
maximum over $\Pi_0$ of the expression (A1) is given by (A4) for any
positive $p_0,p_1$, as can be seen from the proof of Lemma A1, it
follows from Theorem A1 that $p_1 {\rm
tr} \Pi_1 \rho_1 + p_2 {\rm tr}\Pi_2 \rho_2 = 1$ if and only if
$\rho_1 \rho_2 = 0$.  Thus $P_{CM} < 1$.  The contraposition of this
conclusion is the nontrivial part of the theorem. $\Box$

\newpage

\addcontentsline{toc}{section}{Appendix B: Local State Invariance}

\renewcommand{\theequation}{B\arabic{equation}}
\setcounter{equation}{0}

\section*{Appendix B \\
Local State Invariance}

The local state invariance theorem is conceptually significant and has
a simple proof.  \\
\noindent {\em Theorem (Local State Invariance)}:  Let $\rho^{AB}$ be
a state on ${\cal H}^A \otimes {\cal H}^B$ with marginal states $\rho
^A \equiv tr _B \rho^{AB} , \rho ^B$.  The individual or combined
effects of any state transformation and quantum measurement (averaged
over the measurement results) on ${\cal H}^A$ alone leaves $\rho ^B$
invariant.  \\
\noindent{\em Proof}:  It suffices to consider a pure state $|\Phi
\rangle \in {\cal H}^A \otimes {\cal H}^B$ in Schmidt form $|\Phi
\rangle = \sum_k \alpha _k |e_k \rangle |\phi _k \rangle$, $\langle
e_k | e_{k'} \rangle = \langle \phi _k | \phi _{k'} \rangle =
\delta_{kk'}$ so that $\rho ^B = \sum _k |\alpha _k|^2 |\phi _k
\rangle \langle \phi _k |$.  The most general operation on ${\cal
H}^A$ can be represented by extending ${\cal H}^A$ to ${\cal H}^A
\otimes {\cal H}^{A'}$ with initial state $|A'\rangle \in {\cal
H}^{A'}$, and applying a unitary $U$ and measuring a complete
orthonormal basis $\{ |n\rangle \langle n| \}$ on ${\cal H}^A \otimes
{\cal H}^{A'}$ \cite{cpmaps}.  This results in $\tilde{\rho}^B =
\sum_n |n \rangle \langle n | U | \Phi \rangle | A' \rangle \langle A'
| \langle \Phi |U ^{\dagger} |n \rangle \langle n|$ so that $\langle
\phi _k |\tilde{\rho} ^B | \phi _{k'} \rangle = |\alpha _k |^2 \delta
_{kk'} = \langle \phi _k |\rho ^B |\phi _{k'} \rangle$.  The same
result obtains when either $U$ or the measurement on $\{ |n\rangle
\langle n| \}$ is omitted. $\Box$ 
\\
The Schmidt decomposition in the above proof only simplifies
notation and is not essential.  This theorem implies that superluminal
communication via quantum entanglement is impossible, which would be
obtained {\em if and only if} $\rho ^B$ is changed so that a binary
communication channel of classical information with nonzero channel
capacity is created. Observe that the averaging over
measurement results in the theorem is a crucial condition for
application to superluminal communication in which the specific
measurement result on ${\cal H}^A$ is unknown to the party with ${\cal
H}^B$.  While there are many proofs on the impossibility
of entanglement induced superluminal communication in the literature,
see, e.g., \cite{scherer}, none appears to be as complete and simple
as the proof just given.  In particular, the impossibility of cloning
quantum states in some such proofs is not sufficient to establish the
impossibility of superluminal communication.

\newpage

\addcontentsline{toc}{section}{Appendix C: Even and Odd Binomial Sums}

\renewcommand{\theequation}{C\arabic{equation}}
\setcounter{equation}{0}

\section*{Appendix C \\
Even and Odd Binomial Sums}

The even and odd binomial sums used in obtaining (15) are derived as
follows.  Let $P_m$ be the odd sum
\begin{equation}
P_m \equiv \sum^m_{r\,{\rm odd}}\left(
\begin{array}{c}
m \\ r \end{array}\right) p^r (1-p)^{m-r}
\end{equation}
where $p \le \frac{1}{2}$, and let $Q_m$ be the even sum, $Q_m = 1 -
P_m$.  Using the identity
\begin{displaymath}
\left(
\begin{array}{c}
m+1 \\  r\end{array}\right)  = \left(
\begin{array}{c}
m \\ r\end{array} \right)+  \left(
\begin{array}{c}
m \\ r-1\end{array} \right),
\end{displaymath}
the following difference equation for $P_m$ can be derived from (C1):
\begin{equation}
P_{m+1}-P_m = p(Q_m-P_m) = p(1-2 P_m).
\end{equation}
Eq.~(C2) with the initial condition $P_1 = p$ is solved to yield
\begin{equation}
P_m = \frac{1}{2}-\frac{1}{2}(1-2 p)^m.
\end{equation}


\begin{thebibliography}{99}
\bibitem{note_infdim} In the infinite-dimensional case, $U$ is only
a partial isometry \cite{halmos}, but the following condition (32) is still valid.
\end{thebibliography}
\end{document}